----------------------------- Start of body part 2

\documentstyle[11pt]{article}

\begin{document}

\begin{titlepage}
\pagestyle{empty}
\title{On a New Class of Integrable Models}

\author{Alexander Berkovich, C\'esar G\'omez and Germ\'an Sierra \\
Instituto de F\'{\i}sica Fundamental, CSIC \\
Serrano 123.  Madrid, SPAIN}

\maketitle

%\begin{abstract}

%\end{abstract}

\end{titlepage}

\section{Introduction}

\pagestyle{plain}

During the last two years new connections between integrable
models and quantum groups were established. The representation
theory of quantum groups at roots of unit provides new solutions
to the start--triangle equation increasing in this way the
family of lattice integrable models. Most of these new solutions
can be obtained by applying the ``descendent''--procedure [1] to
well known models as for instance the 6--vertex model.

Given an $R$ matrix satisfying the Yang--Baxter equation the first
step in the descendent procedure is to find all possible
solutions to:

\begin{equation}
R (\lambda, \mu) \left[ L (\lambda) \otimes L (\mu) \right]  =
\left[ L (\mu) \otimes L (\lambda) \right] R (\lambda, \mu)
\label{1}
\end{equation}

If the $R$--matrix is, for instance, the quantum $R$ matrix of
$\widehat{SU (2)}_q$ in the 1/2--representation then we should
expect the solutions of (1) to be in one to one correspondence with the
different irreps of the Hopf algebra $\widehat{SU (2)}_q$.
It is in this way that the representation theory plays an
important role in the discovery of new solutions to thestar--triangle equations
by the descendent procedure [2]. The second
step of this method is to find for two given solutions
$L^{\rho_1}, L^{\rho_2}$ of (1) a new $R$--matrix, solution to
the Yang--Baxter equation, and satisfying:

\begin{equation}
{\cal R}^{\rho_1 \rho_2} \left( L^{\rho_1} \otimes L^{\rho_2}
\right) = \left( L^{\rho_2} \otimes L^{\rho_1} \right) {\cal
R}^{\rho_1 \rho_2}
\label{2}
\end{equation}

If we interpret $L^{\rho_1}, L^{\rho_2}$ as associated with two
different irreps $(\rho_1, \rho_2)$ of $\widehat{SU (2)}_q$ the
new ${\cal R}$--matrix will define the intertwiner between these
two representations. The two steps of the descendent--procedure
are graphically represented in fig. 1.

\vspace{4cm}
\begin{center}
Fig. 1. \underline{The descendent--equations}
\end{center}

\vspace{0.3cm}
The simplest example of this procedure is the one known as
fusion where starting, for instance, with the $R^{1/2,
1/2}$--matrix of $\widehat{SU (2)}_q$ we obtain the intertwiner
for regular representations with higher spins [3].

The case of quantum groups at roots of unity is a very
interesting example for the application of this technique. In
this case, the number of different finite dimensional irreps is
much bigger as a consequence of the existence of new central
elements in the Hopf algebra [4]. Hence for the $SU(2)_q$ case, the
central Hopf subalgebra for $q = \varepsilon$ an Nth--root of unit,
contains, in addition to the standard casimir, the new elements
$E^N, F^N, K^N$. The different irreps can be divided into two
main sets: i) regular representations and ii)
generic--representation. The last ones are classified into
cyclic, with non vanishing eigenvalue for $E^N$ and $F^N$,
semicyclic with vanishing eigenvalue only for $E^N$ or $F^N$ but
not both and nilpotent for which $E^N = F^N = 0$ but $K^N$ having
generic values. Starting with the $R$--matrix of $\widehat{SU
(2)}_q$ it is possible to get solutions to the descendent
equations (1) and (2) associated with cyclic, semicyclic and
nilpotent representations. The first case i.e. cyclic, was first
discovered by Bazhanov and Stroganov [1] and the Kyoto group [5]. Surprisingly
enough, the solution obtained coincides with the one
corresponding to the chiral Potts model [6]. These kinds of models are
specially important from a mathematical point of view. Their
spectral manifold is a higher genus curve:

\begin{equation}
\Gamma_k \; : \; x^N + y^N = k (1 + x^N y^N)
\label{3}\end{equation}

\noindent
with $k$ a parameter of the model (for instance $k=0$
corresponds to Fateev--Zamolodchikov model). Given two irreps
$(\xi_1, \xi_2)$ of $\widehat{SU (2)}_q$ with $q^N =1$ and
$\xi \in Spec Z_q$, where $Z_q$ is the central Hopf subalgebra,
then the intertwining condition defines a submanifold in $Spec Z_q$
which can be shown to be isomorphic to the product of two copies
of the spectral curve $\Gamma_k$ of the chiral Potts model.
Moreover the solution $R(\xi_1 \xi_2)$ to equation (2)
factorizes into four pieces which can be represented in terms of
the Boltzman weights $W, \bar{W}$ of the chiral Potts model [5].

The new solution which we would like to discuss in this lecture
corresponds to the case of semicyclic and nilpotent
representations. The model defined in this way shares some
similarities with chiral Potts and with the Heisenberg--Ising models with
higher spin. However, it presents some new features which we now move
on to present.

\section{The semi--cyclic descendent}

In a previous paper [7] we have presented the $R$ matrix solution
to the Yang--Baxter and the intertwiner conditions:

$$
(1 \otimes R (\xi_1 \xi_2)) (R (\xi_1 \xi_3) \otimes 1) (1
\otimes R (\xi_2 \xi_3)) =
\eqno{(3.a)}
$$

\[
= (R (\xi_2 \xi_3) \otimes 1) (1 \otimes R(\xi_1 \xi_3)) (R
(\xi_1 \xi_2) \otimes 1)
\]

$$
R (\xi_1 \xi_2) \Delta_{\xi_1 \xi_2} (a) = \Delta_{\xi_2 \xi_1}
(a) R (\xi_1 \xi_2)
\eqno{(3.b)}
$$

\[
a \in SU (2)_q \; \; \; q^N = 1
\]

\noindent
for $\xi_1, \xi_2$ semicyclic representations of $SU(2)_q$.
Denoting by $x, y, \lambda^N$ the eigenvalues of $E^N, F^N$ and
$K^N$ respectively the semicyclic representations are
parametrized by $(y_1, \lambda_1)$, $(y_2, \lambda_2)$. Using (3.b)
for $a$ in the center, we get the intertwining constraints on
the values of $\xi_1, \xi_2$:
\begin{equation}
\frac{y_1}{1 - \lambda^N_1} = \frac{y_2}{1 - \lambda^N_2} = \chi
\label{4}
\end{equation}

\noindent
with $\chi$ an arbitrary complex number. The solution $R (\xi_1
\xi_2)$ satisfies:

\begin{equation}
\begin{array}{rl}
i) & R (\xi \xi) = 1 \\
ii) & R (\xi_1 \xi_2) R (\xi_2 \xi_1) = 1 \otimes 1 \\
iii) & R (\xi_1 \xi_2) = P R (\xi_2 \xi_1) P
\end{array}
\label{5}
\end{equation}

\noindent
where $P$ is the permutation operator.

For more details on this solution see reference [7].

Following the spirit of the descendent--technology we proceed
now to find solutions to equation (1), for $R$ the quantum
$R$--matrix of $\widehat{SU(2)}_q$ in the 1/2--representation
i.e. the six vertex model, which can be associated with
semicyclic representations of $\widehat{SU(2)}_q$ at $q =
\varepsilon$ an Nth--root of unit (we shall consider $N$ to be
an odd integer). We will first consider the simplest
non trivial case:  $x = y = 0$ and $\lambda$ generic (the
regular representations correspond to $\lambda =
\varepsilon^{2s}$ with $s$ integer or half integer spin).

The 6--vertex $R$--matrix is given by [3]:

\begin{equation}
R^{1/2, 1/2} (u) = sh \left[ u + i\gamma \frac{1}{2} (1 + \sigma^3
\otimes \sigma^3) \right] + i \; {\rm sin\gamma} \; (\sigma^+ \otimes
\sigma^- + \sigma^- \otimes \sigma^+)
\label{6}
\end{equation}

Equation (1) reads now:

\begin{equation}
R^{1/2, 1/2} (u-v) (L (u) \otimes L(v)) = (L (v) \otimes L(u))
R^{1/2, 1/2} (u-v)
\label{7}
\end{equation}

The solutions, we are interested in, are $L^{(\lambda)} (u)$--matrices
acting on $V^{1/2} \otimes V^\lambda$ such that:

\begin{equation}
L^{(\lambda)} (u) : V^{1/2} \otimes V^\lambda \rightarrowV^{1/2} \otimes
V^{\lambda}
\label{8}
\end{equation}

\noindent
where $V^\lambda$ is the irrep of $\widehat{SU
(2)}_\varepsilon$ with eigenvalue of $K^N = \lambda^N$ and where
the tensor product in (7) is defined with respect to the
$V^{1/2}$--indices. Notice that the $L^{(\lambda)} (u)$--matrices we
are looking for define the intertwiners between the irreps 1/2
and $\lambda$. In matrix notation the solution is given by:

$$
L^{(\lambda)} (u) = \frac{1}{e^u \varepsilon^{1/2} \lambda^{1/2}
- e^{-u} \varepsilon^{-1/2} \lambda^{-1/2}} \times
$$

\begin{equation}
\times \left( \begin{array}{cc}
e^u \varepsilon^{1/2} K^{1/2}
- e^{-u} \varepsilon^{-1/2} K^{-1/2} & (\varepsilon -
\varepsilon^{-1}) \varepsilon^{-1/2} FK^{1/2} \\
(\varepsilon- \varepsilon^{-1}) \varepsilon^{-1/2} E K^{-1/2} &
e^u \varepsilon^{1/2} K^{-1/2}
- e^{-u} \varepsilon^{-1/2} K^{1/2} \end{array} \right)
\label{9}
\end{equation}

\noindent
with  $E, K, F$ the generators of $SU(2)_q$ which satisfy the relations:

\[
EK = \varepsilon^2 K E \; \; , \; \; FK = \varepsilon^{-2} KF
\]

\begin{equation}
[E, K] = \frac{K - K^{-1}}{\varepsilon - \varepsilon^{-1}}
\label{10}
\end{equation}

The representation $V^\lambda$ is defined in the basis $\{ |r> \}_{r=0}^{N-1}$

\begin{equation}
\begin{array}{l}
E | r> = d_{r-1} | r-1> \\
F | r> = d_{r} | r+1> \\
K | r> = \lambda \varepsilon^{-2r} | r>
\end{array}
\label{11}
\end{equation}

\[
d^2_j (\lambda) = [j+1] \frac{\lambda \varepsilon^{-j} -
\lambda^{-1} \varepsilon^j}{\varepsilon - \varepsilon^{-1}}
\]
\[
[ x] = \frac{\varepsilon^x - \varepsilon^{-x}}{\varepsilon - \varepsilon^{-1}}
\]

For a graphic representation of the matrix $L^{(\lambda)}(u)$ see
Fig. 2. Notice that each entry in the 2 x 2 matrix (9)
represents an operator acting on the space $V^\lambda$.

\vspace{5cm}
\begin{center}
Fig. 2.  Graphic representation of the $L^{(\lambda)}(u)$--matrix
\end{center}

\vspace{1cm}
The second step in the descendent procedure corresponds to
finding the $R$--matrix solution to the equation (2):

\begin{equation}
{\cal R}^{\lambda_1 \lambda_2} (u-v) \left( L^{(\lambda_1)} (u)
\otimes L^{(\lambda_2)} (v) \right) = \left( L^{(\lambda_2)} (v)
\otimes L^{(\lambda_1)} (u) \right) R^{\lambda_1 \lambda_2} (u-v)
\label{12}
\end{equation}

\noindent
where now the tensor product is defined with respect
to the \underline{$V^{\lambda_1}$ - $V^{\lambda_2}$} \underline{indices} of the
$L^{(\lambda)}$ (u) - matrices (see fig. 1 B). The solution to
(12) defines the intertwiner between the representations
$\lambda_1, \lambda_2$ and it is given by:

\[
R^{\lambda_1 \lambda_2} (u)^{l, r_1 + r_2 - l}_{r_1 r_2} =
\frac{\epsilon^{(r_1 + r_2 - l) l - r_1
r_2}}{\prod^{r_1+r_2-1}_{j=0} (e^u \lambda_1 \lambda_2
\epsilon^{-j} - e^{-u} \epsilon^j)} \times
\]

\[
\times {\sum^{r_1}_{l_1 = 0}
\sum^{r_2}_{l_2 = 0}} \left[ \begin{array}{c} r_1
\\ l_1 \end{array} \right] \left[ \begin{array}{c} r_2 \\ l_2
\end{array} \right] \frac{[l] ! [r_2 - l_2] !}{[r_1 + l_2] !
[r_2]!} (\epsilon - \epsilon^{-1})^{r_1 - l_1 +l_2}
\]

\[
\times \prod^{r_1 + l_2 -1}_{j= r_1} d_j (\lambda_1) \prod^{r_1
+l_2-1}_{j=l_1 +l_2} d_j (\lambda_1) \prod^{r_2
- 1}_{j=r_2 - l_2} d_j (\lambda_2) \prod^{r_1+r_2
-l-1}_{j=r_2 -l_2}
\]

\begin{equation}\times \lambda^{l_2}_1 \lambda^{r_1-l_1}_2 \prod^{r_2 -l_2
-1}_{j=0} (e^u \lambda_2 \epsilon^{-j} - e^{-u} \lambda_1
\epsilon^j) \prod^{l_1 -1}_{j=0} (e^u \lambda_1 \epsilon^{-j+r_2-l_2}
-e^{-u} \lambda_2
\epsilon^{j+l_2 -r_2})
\label{13}
\end{equation}

\noindent
with the following conventions: a) whenever in above products the upper index
is less than the lower index the result is one.
b) the constraint $l_1 + l_2 = l$ must be used to carry out summation.

It is easy to check that this solution coincides in the case of
$u=0$ with the $R^{\xi_1 \xi_2}$--matrix of reference [7] for
$\xi_1, \xi_2$ two semi--cyclic irreps with $x=y=0$ and $K^N =
\lambda^N_1, \lambda^N_2$. Moreover  for $\lambda_1 = \lambda_2
= \epsilon^{2s}$ eqn. (13) gives us the spin $s - R$ matrix.

Summarizing, we have obtained starting with the $R$--matrix of
the six vertex model for $q$ an Nth-root of unit a class of
descendent models characterized by the quantum $R$--matrix (13).
The transfer matrix of these models can be defined for periodic
boundary conditions as follows (see fig. 3):

\begin{equation}
T_{\lambda_0} (u, \lambda) : \otimes^L V^{\lambda_0}
\rightarrow
\otimes^L V^{\lambda_0}
\label{14}
\end{equation}

\begin{equation}
\langle r'_1 \cdots r'_N | T_{\lambda_0} (u, \lambda) | r_1,
r_2 \cdots r_N \rangle = \left( \sum_{l's} R^{\lambda,
\lambda_0} (u)_{l_1 r_1}^{l_2 r'_1} R^{\lambda \lambda_0}
(u)^{l_3 r'_2}_{l_2 r_2} \cdots \right)
\label{15}
\end{equation}

\noindent
where $\lambda_0$ appears as a parameter characterizing the model
and $u$ and $\lambda$ as spectral variables. The Yang-Baxter
relation for the $R$--matrix (eq. 13) implies the integrability equation:

\begin{equation}
\left[ T_{\lambda_0} (u, \lambda'), T_{\lambda_0} (u', \lambda'') \right]
= 0
\label{16}
\end{equation}

Equation (16) is the standard integrability condition for
soluble models with the important peculiarity that the spectral
variables are now living on a manifold of complex dimension
equal 2. Notice that for the kind of irreps we are considering
(i.e. with vanishing eigenvalues of $E^N$ and $F^N$ and generic
eigenvalue of K) the intertwining condition is not imposing any
constraint on the allowed values of $\lambda$.

\[(\lambda'_1 u) \begin{array}{cc|cc|cc|cc|cc|ccc}
& & r'_1 & & r'_2 & & r'_3 & & r'_4 \cdots & & r'_L & & \\
\hline
l_1 & & &l_2  & & l_3  & & l_4  & & l_5 \cdots  & e 1  & \\
& & r_1  & & r_2  & &  r_3 & & r_4 \cdots & & r_L & | r_1 \cdots
r_L> \\
\end{array}
\overbrace{\otimes^L V^{\lambda_0}}
\]

\begin{center}
$L = \#$ sites in the row.

Fig. 3: The transfer matrix $T_{\lambda_0} (\lambda, u)$
\end{center}

\noindent

\section*{3. The associated 1D--chain}

Given the transfer matrix (14-15) we can define a local 1D
hamiltonian as follows:

\begin{equation}
\left. H^{(\lambda_0)} = i \frac{\partial}{\partial u}  l n
T_{\lambda_0} (u, \lambda_0) \right|_{u= 0}
\label{17}
\end{equation}

\noindent
which will be hermitian for certain allowed regions of
$\lambda_0$. This hamiltonian defines a one dimensional
spin chain with the Hilbert space given by:

\begin{equation}
{\cal H} = \otimes^L V^{\lambda_0}
\label{18}
\end{equation}

The most natural interpretation of this spin--chain is as a
generalization to continuous ``spin" of the higher spin quantum
chains defined in references [8]. In fact, as we have
already mentioned for the special case $\lambda_0 =
\epsilon^{2S_{max}} (2S_{max} +1 = N)$ the hamiltonian (16)
coincides with the hamiltonian $H^{S_{max}}_\epsilon$ of spin
$S_{max}$ and anisotropy depending of $\epsilon$.

The two main new features of the model defined by the transfer
matrix (14) are:

\begin{itemize}
\item[i)] All the $V^{\lambda_0}$ representations are of the
same dimensions $N$. This is specific of working with the
quantum group at roots of unit.
\item[ii)] The integrability condition (16) with two independent
spectral parameters.
\end{itemize}

The main consequence of (16) for the spin chain defined by (16)
is the existence of a local operator $Q^{(\lambda_0)}$ defined by:

\begin{equation}
Q^{(\lambda_0)} = \left. 2 \lambda \frac{\partial}{\partial \lambda} l n
T_{\lambda_0} (u = 0,\lambda) \right|_{\lambda = \lambda_0}
\label{19}
\end{equation}

\noindent
such that:

\begin{equation}
\left[ H^{\lambda_0}, Q^{\lambda_0} \right] = 0
\label{20}
\end{equation}

Denoting $\xi \equiv (u, \lambda)$, the transfer matrix (14) can
be written as:

\begin{equation}
T_{\xi, \xi'} = T_{(u, \lambda)|(u'\lambda_0)} \equiv
T_{\lambda_0} (u - u', \lambda)
\label{21}
\end{equation}

\noindent
where we have used the difference property with respect to the
spectral variable $u$. The most general local hamiltonian we can
define is:

\begin{equation}
{\cal H} = \frac{d}{dt} \left. ln T_{\xi, \xi'(t)}
\right|_{t= 0}
\label{22}
\end{equation}

\noindent
with $\lim_{t
\rightarrow 0} \xi'(t) =  \xi$. This definition of the hamiltonian would
be, in spirit, very close to the way a 1D hamiltonian is defined
for the chiral Potts model. However, it is clear that in our case
any Hamiltonian obtained as indicated by eqn. (22) will be a
linear combination of $H^{(\lambda_0)}$ and $Q^{(\lambda_0)}$.
Before entering into the detailed study of the 1D hamiltonians, it
is worthwhile to discuss the differences between the models we
are defining and the chiral Potts. The model which we can compare
with is the Fateev--Zamolodchikov model. The way the cyclicity
of the irreps, entering into the definition of this model, is reflected
in the physics is through the $Z(N)$--invariance. In fact, the spectrum of
the hamiltonian decomposes in different sectors with well
defined $Z(N)$--charge.

In the noncyclic case, we are presenting here, there exists a well
defined ``reference state" $| \Omega_0 > \in \otimes^L V^{\lambda_0}$:

\begin{equation}
| \Omega_0 > = | 0,0 \cdots 0 >
\label{23}
\end{equation}

Recall that the states in $\otimes^L V^\lambda$ are given by $|r_1
r_2 \cdots r_L>$ with $r_i = 0,\cdots,N-1$. The reference state
is the natural generalization of the ferromagnetic state with
all the spins up. The conserved number, as in the case of the
Heisenberg chains, is the total number of spins ``down'' which in our
case is given by:

\begin{equation}
\sum^L_{i=1} r_i = \rho.
\label{24}
\end{equation}

Different sectors correspond to different values $\rho$. In the
chiral Potts model, i.e. cyclic case, there is no good
definition of ``reference state" and the conserved quantum
number is given by the $Z(N)$ charge defined by:

\begin{equation}
X = e^{2 i\pi Q/N}.
\label{25}
\end{equation}

\noindent
where

\begin{equation}
X | r_1 \cdots r_L >= |r_1 +1 \cdots r_2 +1>.
\label{26}
\end{equation}

In the next section and using the fact that we have a good
reference state we will proceed to diagonalize the hamiltonian
$H^{\lambda_0}$ by using the Bethe Ansatz technique.

\section*{4. Bethe Ansatz Equations}

In terms of the $L^{\lambda_0} (u)$--matrices we define the
monodromy matrix by:

\begin{equation}
t_{\lambda_0} (u) = L^{\lambda_0}_L (u) \cdots L^{\lambda_0}_2
L^{\lambda_0}_1 (u)\label{27}
\end{equation}

\noindent
with $L$ the length of the row in lattice units. The monodromy
matrix $t_{\lambda_0}(u)$ satisfies:

\begin{equation}
R^{1/2, 1/2} (u - v) t_{\lambda_0} (u) t_{\lambda_0} (v)
= t_{\lambda_0} (v) t_{\lambda_0} (u) R^{1/2, 1/2} (u - v)
\label{28}
\end{equation}

Representing it as a 2x2 matrix:

\begin{equation}
t_{\lambda_0} (u) = \left[ \begin{array}{cc} A_{\lambda_0} (u) &
B_{\lambda_0} (u) \\
C_{\lambda_0} (u) & D_{\lambda_0} (u) \end{array} \right]
\label{29}
\end{equation}

\noindent
we obtain the operators $B_{\lambda_0} (u) : \otimes^L
V^{\lambda_0} \rightarrow \otimes^L V^{\lambda_0}$ which can be
interpreted as creating on the reference state (23) an
``elementary excitation". Notice that:

\begin{equation}
B_{\lambda_0} (u_1) B_{\lambda_0} (u_2) = B_{\lambda_0} (u_2)
B_{\lambda_0} (u_1)
\label{30}
\end{equation}

$$
C_{\lambda_0} (u) | \Omega_0 > = 0.
\eqno{(30;b)}
$$

To diagonalize the transfer matrix (13) we use the standard
algebraic Bethe Ansatz:

\begin{equation}
|\psi> = \prod^M_{i=1} B_{\lambda_0} (u_i) | \Omega_0 >
\label{31}
\end{equation}

The number $M$ of ``excitations" is for our model a conserved
quantum number and therefore we can diagonalize the transfer
matrix in each sector.

Using equation (9) it is easy to find the eigenvalues $\Lambda
(\lambda_0, \lambda, u; \{ u_i \}_{i=1, \cdots, M})$ of the
transfer matrix T in each sector:

\begin{equation}T_{\lambda_0} (u, \lambda) |\psi> = \Lambda (\lambda_0,
\lambda,
u; \{u_i \}) |\psi > +\;unwanted\;terms.
\label{32}
\end{equation}

\noindent
with

\begin{equation}
\Lambda (\lambda_0, \lambda,
u; \{u_i \}) = \sum^{N-1}_{r=0} (R^{r0}_{r0} (\lambda,
\lambda', u))^L \prod^M_{i=1} {\cal L}^{\lambda'}_r (u - u_i)
\label{33}
\end{equation}

\noindent
where the $R$ matrix is the one given in (12) and
${\cal L}^{\lambda'}_r (u-u_i)$ is:

\begin{equation}
{\cal L}^{\lambda'}_r (u)= \frac{(\epsilon^u \epsilon^{1/2}
\lambda^{\prime 1/2} - e^{-u} \epsilon^{-1/2} \lambda'^{-1/2}) (e^u
\epsilon^{-1/2} \lambda'^{-1/2} - e^{-u} \epsilon^{1/2}
\lambda'^{1/2})}{(e^u \epsilon^{1/2 - r} \lambda'^{1/2} - e^{-u}
\epsilon^{r-1/2} \lambda'^{-1/2}) (e^u \epsilon^{-r-1/2}
\lambda'^{1/2} - e^{-u} \epsilon^{r+1/2} \lambda'^{-1/2})}
\label{34}
\end{equation}

To fix the values of the ``rapidities'' $u_i$ we must eliminate
the unwanted terms in (32). This can be done using equation (28)
or more easily imposing that the residues at the poles both in
$u$ and $\lambda$ of (33) vanish. The result is:

\begin{equation}
\left( \frac{e^{u_j} \epsilon^{1/2} \lambda^{1/2}_0 - e^{-u_j}
\epsilon^{-1/2} \lambda^{-1/2}_0}{e^{u_j} \epsilon^{1/2}
\lambda^{-1/2}_0 - e^{-u_j} \epsilon^{-1/2} \lambda^{1/2}_0}
\right)^L = \prod^{M}_{\begin{array}{c} k=1 \\ k \neq j
\end{array}} \frac{sh (u_j - u_k + i \gamma)}{sh (u_j - u_k - i \gamma)}
\label{35}
\end{equation}

\noindent
where $\epsilon = e^{i \gamma}$. In the case $\lambda_0 =
\epsilon^{2s}$ with 2s being an integer, i.e. regular representations of spin
s, equations (35) become the Bethe Ansatz equations for the higher spin
Heisenberg--Ising chains of references [8].

{}From (33) we can find the eigenvalues of the operators
$H^{\lambda_0}$ and $Q^{\lambda_0}$ defined in the previous section:

\begin{equation}
E^{\lambda_0} = 2i \sum^M_{j=1} \frac{\lambda_0 -
\lambda^{-1}_0}{(e^{u_j} \varepsilon^{1/2} \lambda^{1/2}_0 -e^{-u_j}
\varepsilon^{-1/2} \lambda^{-1/2}_0 ) (e^{u_j}
\varepsilon^{1/2} \lambda^{-1/2}_0 - e^{-u_j} \varepsilon^{-1/2}
\lambda^{1/2}_0)}
\label{36}
\end{equation}

\begin{equation}
Q^{\lambda_0} = 2 \sum^M_{j=1}
\frac{e^{2u_j} \varepsilon - e^{-2u_j} \varepsilon^{-1}}{(e^{u_j}
\varepsilon^{1/2} \lambda^{1/2}_0 -
e^{-u_j} \varepsilon^{-1/2} \lambda^{-1/2}_0 ) (e^{u_j}
\varepsilon^{1/2} \lambda^{-1/2}_0 - e^{-u_j} \varepsilon^{-1/2}
\lambda^{1/2}_0)}
\label{37}
\end{equation}

For the intermediate case corresponding to $\lambda_0 =
\varepsilon^{2s}$ with arbitrary $s$, the equations (36) (37) become:

\begin{equation}
E^s = - \sum^M_{j=1} \frac{\sin (2 \gamma s)}{sh [
\frac{\gamma}{2} (\alpha_j + 2 is)] sh [ \frac{\gamma}{2} (\alpha_j -
2 is)]}
\label{38}
\end{equation}

\begin{equation}
Q^s =  \sum^M_{j=1} \frac{s h ( \gamma \alpha_j)}{sh [
\frac{\gamma}{2} (\alpha_j + 2 is)] sh [ \frac{\gamma}{2} (d_j -
2 is)]}
\label{39}
\end{equation}

\noindent
where we have introduced the new variable $\alpha$ defined by

\begin{equation}
u + \frac{i}{2} \gamma = \frac{\gamma}{2} \alpha
\label{40}
\end{equation}

Now we briefly discuss the case of the nonvanishing $\chi$, defined by eqn.(4).
The first observation to be made is that the number of excitations
``$M$'', introduced in (31), is no longer ``good quantum number".
What replaces it is an eigenvalue of $\Delta K$ (which is $\lambda^L_0
e^{i\frac{2 \pi Q}{N}}$ for $Q=0, 1, \cdots N-1)$ since this
operator commutes with both transfer matrices, namely

\[
\Delta K = K^{\lambda_0}_L \cdots K_2^{\lambda_0} K_1^{\lambda_0}
\]

\begin{equation}[ \Delta K, tr t_{\lambda_0} (u)] = [\Delta K, T_{\lambda_0}
(u,
\lambda) ] =0
\label{41}
\end{equation}

Surprisingly enough, some nice features of $\chi =0$ case
survive. In particular inspecting formula (13), one concludes
that R--matrix preserves its low--triangular form while acting
on the ``reference state" (23). Therefore, this state remains an
eigenvector of $T_{\lambda_0} (u_1 \lambda)$. More precisely

\begin{equation}
T_{\lambda_0} (u, \lambda) | \Omega_0 > = \sum^{N-1}_{r=0} (R^{r0}_{r0}
(\lambda_0, \lambda, u))^L | \Omega_0 >
\label{42}
\end{equation}

\noindent
What makes $\chi \neq 0$ situation somewhat more complicated is
the fact that the commutation relations for $A_{\lambda_0},
B_{\lambda_0}, C_{\lambda_0}, D_{\lambda_0}$ and the elements of
$T_{\lambda_0}(u,\lambda)$ no longer have simple form and
therefore, one does not expect for the state (31) to be an
eigenvector of $T_{\lambda_0} (u, \lambda)$. Consequently,
the appropriate generalization of Algebraic Bethe Ansatz is
called for in this case.

To get the feeling of what transpires, one should note that the
state (31) is still an eigenvector of $tr t_{\lambda_0} (u)$

\begin{equation}
tr t_{\lambda_0} (u) | \psi_M > = \sum^1_{r=0} \left( L^{\lambda_0}
(u)^{r0}_{r0} \right)^L
\prod^M_{i=1} R^{1/2, 1/2} ( (-1)^r (u_i - u) )^{01}_{10}
| \psi_M >
\label{43}
\end{equation}

Indeed, in deriving eqn.(43) above, one only uses 6--vertex
commutation relations (28) and the degeneracy eqn.(30;b) for the
operator $C_{\lambda_0} (u)$. Since neither eqn.(28) nor eqn.(30;b)
are affected by turning on nonzero $\chi$, one could infer that
eqn.(43) should hold true. Thus, it appears that going from $\chi
=0$ to $\chi \neq0$ case produces no visible effect on the
spectrum of the transfer matrix $tr t_{\lambda_0} (u)$. This
conclusion, however, may be a bit premature. To see that, let us
recall, that not all the solutions of (35), generally speaking,
correspond to non-zero vectors of the form (31).

Quite frequently an additional investigation (involving delicate
limiting procedures) is required to determine the fate of the
particular solution. In general, the answer to {\it ``To Be Or
Not To Be}" question may depend on whether or not $\chi$ takes
on the zero value.
Let us now generate \underline{finite--dimensional} vector space
$V_M (Q = M - int [ \frac{M}{N}])$ by premultiplying non-zero
vector $| \psi_M >$ by any sum of products of operators
$T_{\lambda_0} (u, \lambda), T_{\lambda_0} (u', \lambda')
\cdots$ for any $u, \lambda; u', \lambda'; \cdots$ (but all
with the same $\lambda_0$ and $\chi$). It is clear, that in this
space one can diagonalize simultaneously the whole family of
commuting transfer--matrices $T_{\lambda_0} (u, \lambda)$.
Recalling that $tr t_{\lambda_0} (u)$ commutes with
$T_{\lambda_0} (u, \lambda)$, one immediately arrives to the conclusion
that the eigenvectors of family $T_{\lambda_0} (u, \lambda)$
can be constructed as linear combinations of vectors (31), all
of the same length (modulo N) and eigenvalue of transfer--matrix
$tr t_{\lambda_0} (u)$. Apparently, what seems to be going on is
as follows: The eigenvalues of transfer matrices $T_{\lambda_0}
(u, \lambda)$ and  $tr t_{\lambda_0} (u)$ are highly degenerate
when $\chi = 0$. Turning on finite $\chi$ results in lifting this
degeneracy for $T_{\lambda_0} (u, \lambda)$ and changing
multiplicities of eigenvalues of $tr t_{\lambda_0} (u)$. This is
very intriguing phenomena and certainly warrants further
investigations. To summarize, we can still characterize
eigenstates of commuting family $T_{\lambda_0} (u, \lambda)$
by the set of Bethe Ansatz roots (35), abandoning, however, the
simple representation for Bethe vector given by formula (31). The
generalization of algebraic Bethe Ansatz, briefly sketched
above, is similar in spirit to Tarasov's proposal for
superintegrable chiral Potts Model [9].

This similarity is, by no means, accidental. Indeed, chiral Potts
model on superintegrable line shares an important property of
semi--cyclic case: The L--matrix which intertwines spin 1/2 and
cyclic repr. has ``top" reference state but no bottom state.
Whether this similitude has a deeper implication remains to be seen.

When algebraic Bethe Ansatz does not work (or does not provide
the shortest route), one may resort to the alternative procedure
in order to solve for eigenvalues of Hamiltonian and transfer
matrix. This procedure, known as ``Functional Relations"
method, was exploited quite successfully in recent years to
find the spectrum of RSOS [10] and chiral Potts models [11].
The gist of this approach can be described in a few words as
follows: One keeps on fusing various R--matrices (related to the
model under investigation) throwing in some obvious (and not so
obvious) symmetries along the way, until one comes back to where
journey began. The result is the system of functional equations
for the transfer matrices. In a sense, one can regard this
procedure as a generalization of Zamolodchikov--Karowski
Bootstrap program [12] to determine S-matrices of exactly
integrable models.

In order to apply this technique to our model, let us recall
that the tensor product of a spin $j$ representation with a
semi--cyclic one is completely reducible. In particular,
for $j= \frac{1}{2}$, we have (see reference [14])
\begin{equation}
( \frac{1}{2} ) \otimes (y, \lambda) = ( y, \epsilon \lambda)_+
+ (y, \epsilon^{-1} \lambda)_-
\label{44}
\end{equation}

In each subspace one can find the highest weight vector $V^+_0$
and $V^-_0$ such that

\begin{equation}
\Delta(E) V^{\pm}_0 = 0 \;\;and\;\;\Delta (K) V^\pm_0 =
\epsilon^{\pm1} \lambda V^\pm_0
\label{45}
\end{equation}

Making use of (11) we obtain for each sign a semi--cyclic
representation with the basis $[ V^\pm_0 , \cdots, V^\pm_{N-1} ].$

Let us now define two projection points $u_\pm$ as follows:

\begin{equation}
L^{\lambda_0} (u_\mp) V^\pm_i = 0
\label{46}
\end{equation}

At these points the operators $L^{\lambda_0}$ becomes
essentially the projector $P^\pm$ onto corresponding $\pm$ subspace.

\begin{equation}
L^{\lambda_0} (u_{\pm}) \sim P^\pm;\;\;P^+ + P^- = 1\;\;and\;\;P^+P^-
= P^- P^+ = 0
\label{47}
\end{equation}

It follows then from eqn. (12) that

\[
P^- [ L^{\lambda_0} (u) \otimes R^{\lambda_1, \lambda_0} (u
- u_-)] = \cdots [ R^{\lambda_1, \lambda_0} (u - u_-) \otimes
L^{\lambda_0} (u)] \cdots P^-
\]

\noindent
and

\begin{equation}
P^- [ L^{\lambda_0} (u) \otimes R^{\lambda_1, \lambda_0} (u - u_-)
] P^+ = 0
\label{48}
\end{equation}

The relation (48) above reveals a block--triangular structure of
the product

\[O^{-1} [ L^{\lambda_0} (u) \otimes R^{\lambda_1, \lambda_0} (u -
u_-) ] O =
\left(
\begin{array}{cc}
P^+ LRP^+ & * \\
0 & P^-LRP^-
\end{array}
\right)
\]

Remarkably, $P^+LRP^+$ and $P^-LRP^-$ turn out to be

\[
P^+ L^{\lambda_0} (u) \otimes R^{\lambda_1, \lambda_0} (u - u_-)
P^+ =
\]

\begin{equation}
= L^{\lambda_0} (u)^{00}_{00} R^{\epsilon \lambda_1, \lambda_0}
(u - u_- + i \frac{2\pi}{N})
\label{49}
\end{equation}

\[
P^- L^{\lambda_0} (u) \otimes R^{\lambda_1, \lambda_0} (u - u_-)
P^- =
\]
\begin{equation}
= L^{\lambda_0} (u)^{10}_{10} R^{\epsilon^{-1} \lambda_1,
\lambda_0} (u - u_- - i \frac{2\pi}{N})
\label{50}
\end{equation}

In the product of block--triangular matrices the diagonal blocks
are multiplied independently. Thus, we have for transfer--matrices

\[
tr t_{\lambda_0} (u) T_{\lambda_0} (u - u_-, \lambda_1) =
[L^{\lambda_0} (u)^{00}_{00}]^L T_{\lambda_0} (u-u_- + i
\frac{2\pi}{N}, \epsilon^{\lambda_1}) +
\]

\begin{equation}
+ (L^{\lambda_0} (u)^{10}_{10} )^L T_{\lambda_0} (u - u_- - i
\frac{2\pi}{N}, \epsilon^{-1} \lambda_1)
\label{51}
\end{equation}

The use of the second projection point $u_+$ leads to the
similar equation. Note, that $T_{\lambda_0} (u, \lambda)$
depends essentially on two parameters $u$ and $\lambda$.
Therefore, another functional relation is needed in order to
determine eigenvalues of $T_{\lambda_0} (u, \lambda)$ interms of known
eigenvalues of $tr t_{\lambda_0} (u)$. This
additional relation (along with eigenvalues of $T_{\lambda_0}
(u, \lambda))$ will be presented in the forthcoming publication [13].

\end{document}